\documentclass[a4paper,11pt]{article}
\usepackage{pos}

\newcommand{\GeV}{{\ensuremath\rm GeV}}
\newcommand{\TeV}{{\ensuremath\rm TeV}}
\newcommand{\pb}{{\ensuremath\rm pb}}
\newcommand{\fb}{{\ensuremath\rm fb}}
\newcommand{\lb}{\left(}
\newcommand{\rb}{\right)}
\newcommand{\al}{\alpha}

\title{TRSM benchmark planes - {EPS-HEP2023 update}}
\ShortTitle{TRSM benchmark planes}

\author*[a]{Tania Robens}

\affiliation[a]{Rudjer Boskovic Institute,\\
  Bijenicka cesta 54, Zagreb, Croatia}

\emailAdd{trobens@irb.hr}

\abstract{ I briefly review the Benchmark Planes in the Two-Real-Singlet Model (TRSM), a model that enhances the Standard Model (SM) scalar sector by two real singlets that obey a $\mathbb{Z}_2\,\times\,\mathbb{Z}_2'$ symmetry. All fields acquire a vacuum expectation value, that softly breaks the above symmetry and leads to mixing of all CP even neutral states. Furthermore,  all interactions with SM-like particles are inherited from the SM-like doublet via mixing. I remind the readers of the previously proposed benchmark planes, and briefly discuss possible production at future Higgs factories, as well as regions in a more generic scan of the model. Results are based on previous work presented in \cite{Robens:2019kga,Robens:2022nnw,Robens:2023pax}.\\
{RBI-ThPhys-2023-43}}

\FullConference{The European Physical Society Conference on High Energy Physics (EPS-HEP2023)\\
 21-25 August 2023\\
Hamburg, Germany\\}

\begin{document}
\maketitle

\section{Introduction and benchmark planes}
The model presented here has already been widely discussed in the literature \cite{Robens:2019kga,Papaefstathiou:2020lyp,Robens:2022nnw}, to which we refer the reader for more details. We here just briefly repeat the main characteristics for reference. The Two-Real-Singlet Model (TRSM) is a new physics model that enhances the Standard Model (SM) electroweak sector by two additional fields that are singlets under the SM gauge group. The fields obey an additional $\mathbb{Z}_2\,\otimes\,\mathbb{Z}_2'$ symmetry.  All scalar fields acquire a vacuum expectation value, and all CP even neutral states are related to the gauge eigenstates via mixing. In order to comply with current observations by the LHC experiments, one of the resulting three CP even neutral scalars has to show properties in agreement with the measurements of the Higgs boson by the LHC experiments \cite{ATLAS:2022vkf,CMS:2022dwd}.

The original benchmark planes (BPs) have been proposed in \cite{Robens:2019kga}. They can be classified as
either asymmetric (AS) production and decay, in the form of
$p\,p\,\rightarrow\,h_3\,\rightarrow\,h_1\,h_2$,
or
symmetric (S) decays in the form of
$p\,p\,\rightarrow\,h_i\,\rightarrow\,h_j\,h_j$,
where in our study none of the scalars corresponds to the 125 \GeV~ resonance.  We follow the convention that $M_1\,\leq\,M_2\,\leq\,M_3$ for the masses of the scalars $h_{1,2,3}$ in the mass eigenstates. 

The BPs have been widely discussed in e.g. \cite{Robens:2019kga, Robens:2022nnw, Robens:2023pax}. Although by now several searches exist that in principle are sensitive to some of the benchmark planes (see section \ref{sec:lhc}), the model still renders quite a few final state signatures that so far have not yet been investigated by the LHC experiments. We here display four of the BPs, where we also list maximal cross sections that can be obtained at a 13 \TeV~ $pp$ collider:

\begin{itemize}
 \item{} {\bf   AS} {\bf   BP2: $h_3 \to h_1 h_2$ ($h_2 = h_{125}$)}: {SM-like decays for both scalars: $\sim\,0.6\,\pb$
\item {\bf   AS} {\bf   BP3: $h_3 \to h_1 h_2$ ($h_1 = h_{125}$)}: { (a) SM-like decays for both scalars $\sim\,0.3\,\pb$; {  (b) $h_1^3$ final states: $\sim\,0.14\,\pb$}}
  \item {\bf   S} {}{\bf   BP4: $h_2 \to h_1 h_1$ ($h_3 = h_{125}$)}: {up to 60 \pb}
 \item {} {\bf   S} {\bf   BP6: $h_3 \to h_2 h_2$ ($h_1 = h_{125}$)}:
SM-like decays: up to 0.5 \pb; {$h_1^4$ final states: around 14 \fb}}
\end{itemize}

\begin{center}
\begin{figure}[tbh]
\begin{center}
\includegraphics[width=0.45\textwidth]{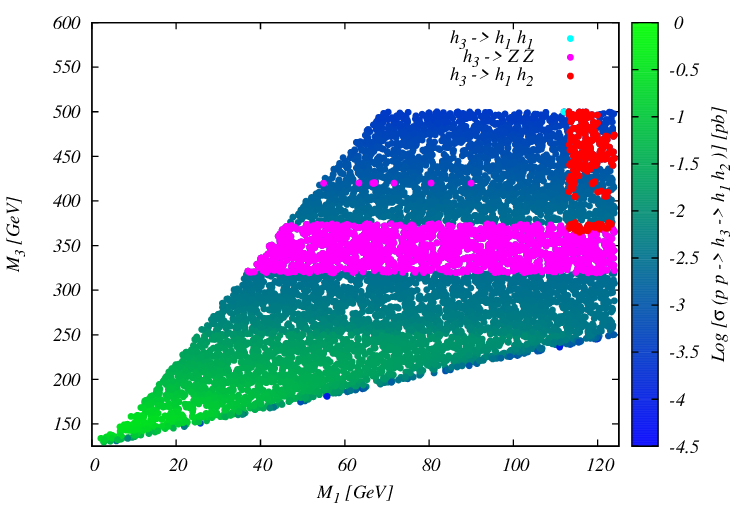}
\includegraphics[width=0.45\textwidth]{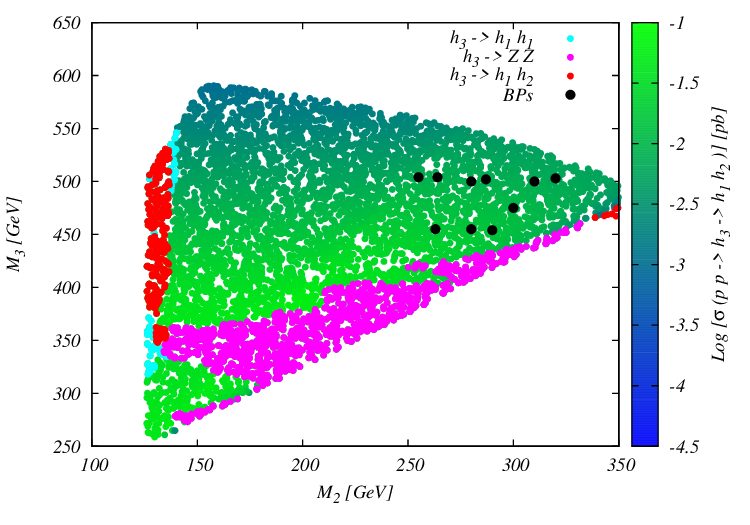}
\caption{\label{fig:bp2bp3} Parameter regions in BP2 {\sl(left)} and BP3 {\sl (right)} with updated constraints; see text for details.}
\end{center}
\end{figure}
\end{center}

\begin{center}
\begin{figure}[tbh]
\begin{center}
\includegraphics[width=0.45\textwidth]{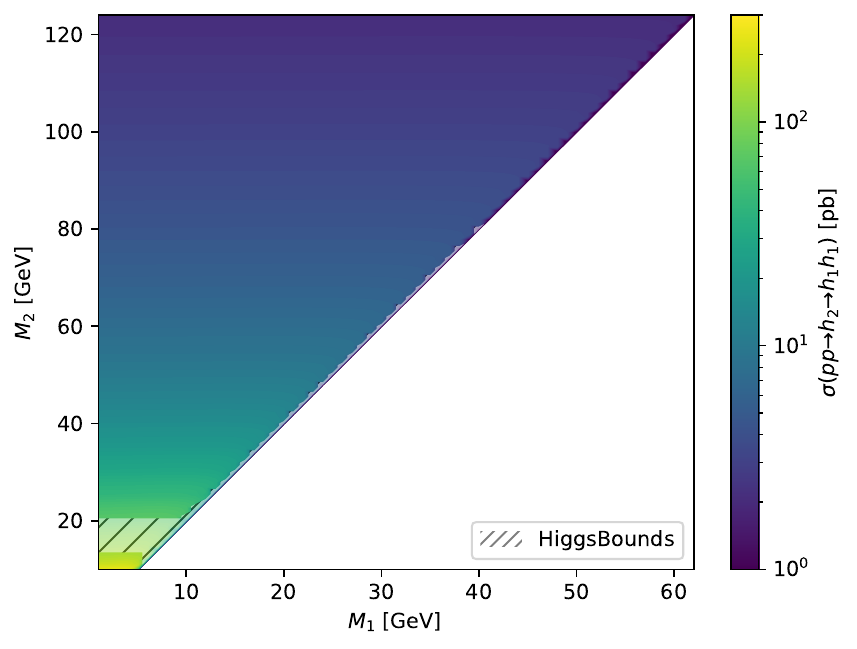}
\includegraphics[width=0.45\textwidth]{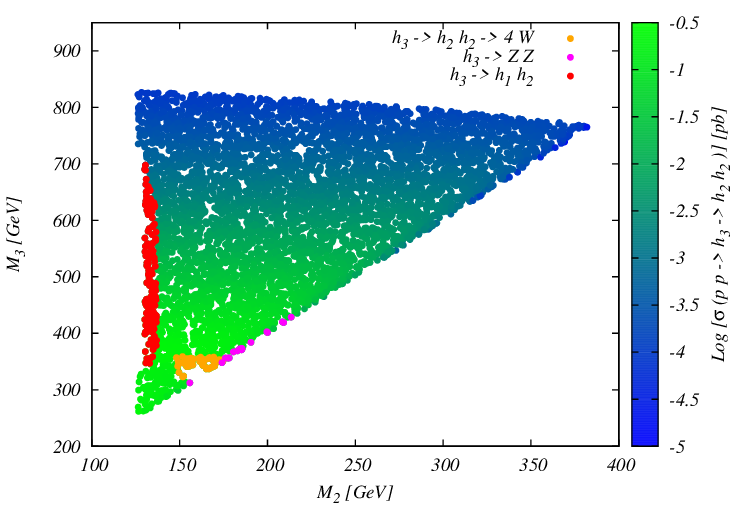}
\caption{\label{fig:bp4bp6} Parameter regions in BP4 {\sl(left)} and BP6 {\sl (right)} with updated constraints; see text for details.}
\end{center}
\end{figure}
\end{center}

The corresponding BPs are shown in figures \ref{fig:bp2bp3} and \ref{fig:bp4bp6}, respectively. The label $h_3\,\rightarrow\,h_1\,h_2$ refers to regions in the parameter space excluded by \cite{CMS:2021yci}, while $h_3\,\rightarrow\,h_1\,h_1$ correspond to results from \cite{CMS:2018ipl,ATLAS:2018rnh,CMS:2018qmt}  and $h_3\,\rightarrow\,Z\,Z$ from \cite{ATLAS:2020tlo}, respectively. For BP6, the ATLAS search $h_3\,\rightarrow\,h_2\,h_2\,\rightarrow\,W^+\,W^-\,W^+\,W^-$ \cite{ATLAS:2018ili} is also sensitive. A more detailed discussion on updates is presented in \cite{Robens:2023pax}. The exclusion bounds have been obtained via the publicly available tool HiggsTools \cite{Bahl:2022igd}.

\section{LHC interpretations}\label{sec:lhc}

\begin{center}
\begin{figure}[tbh]
\begin{center}
\includegraphics[width=0.9\textwidth]{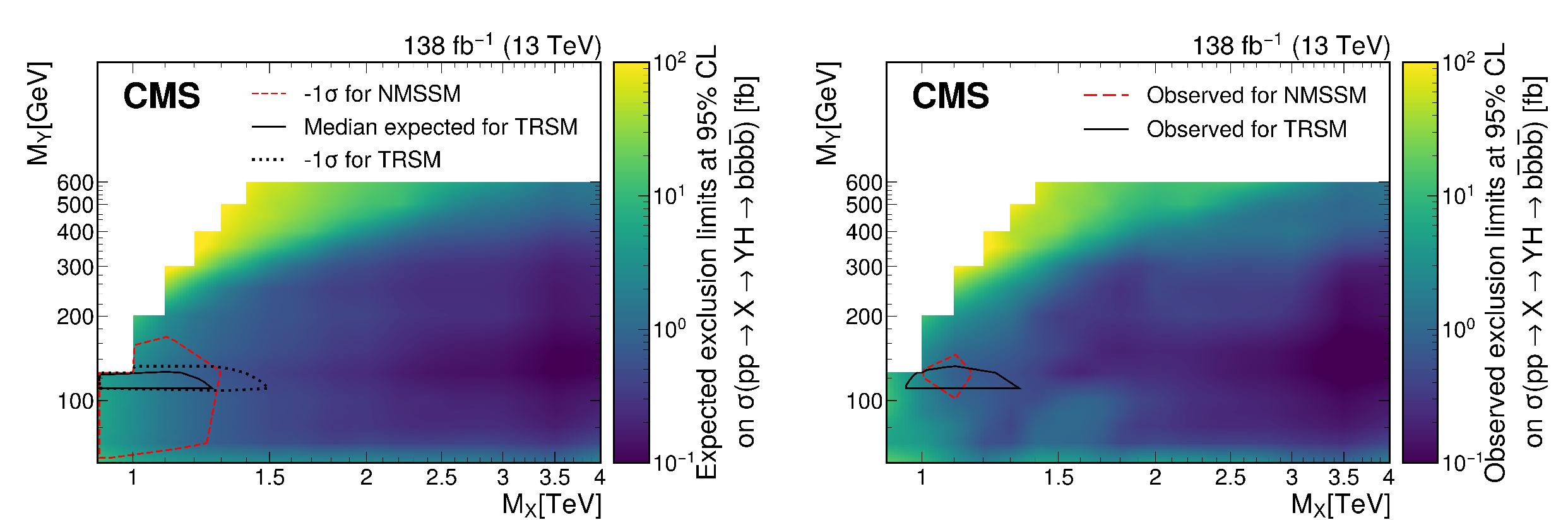}
\caption{\label{fig:cmsres} Expected {\sl (left)} and observed {\sl (right)} $95\%$ confidence limits for the $p\,p\,\rightarrow\,h_3\,\rightarrow\,h_2\,h_1$ search, with subsequent decays into $b\bar{b}b\bar{b}$. For both models, maximal mass regions up to $M_3\,\sim\,\,1.4\,\TeV,\;M_2\,\sim\,\,140\,\GeV$ can be excluded.}
\end{center}
\end{figure}
\end{center}

\begin{center}
\begin{figure}[tbh]
\begin{center}
\includegraphics[width=0.45\textwidth]{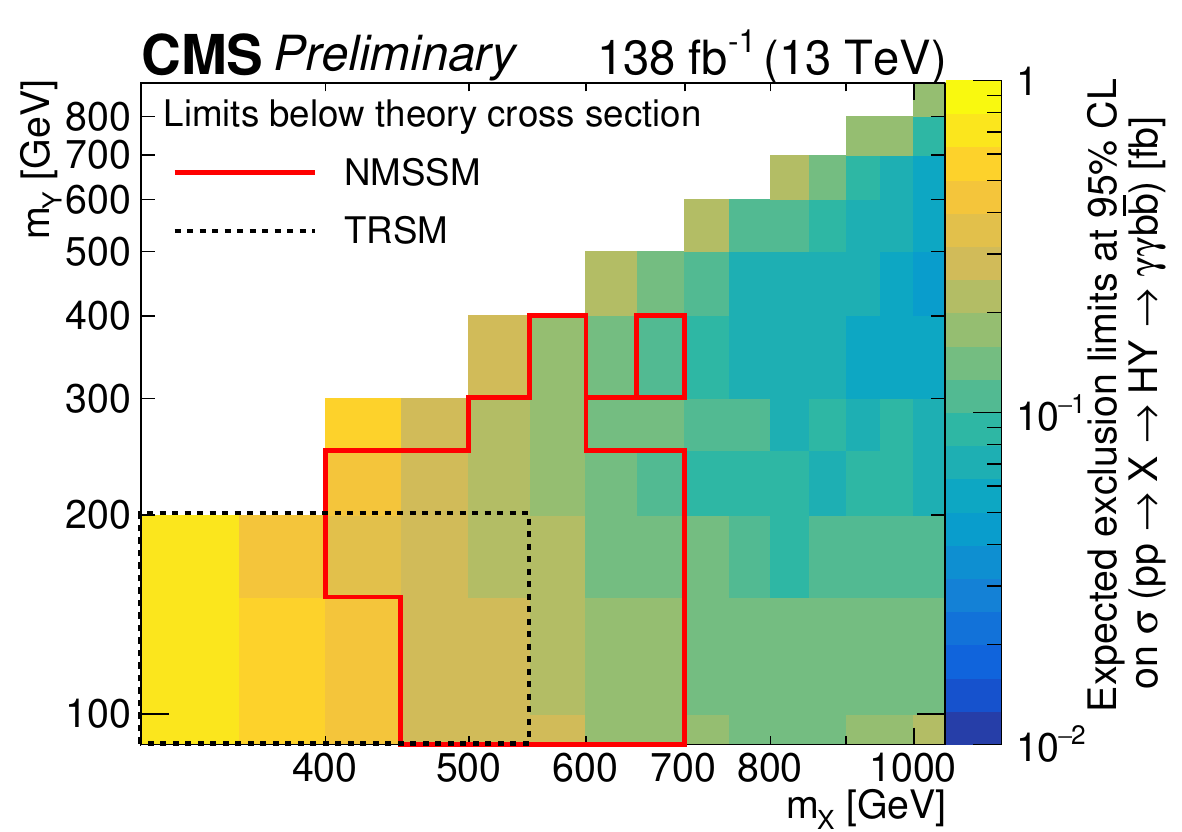}
\includegraphics[width=0.45\textwidth]{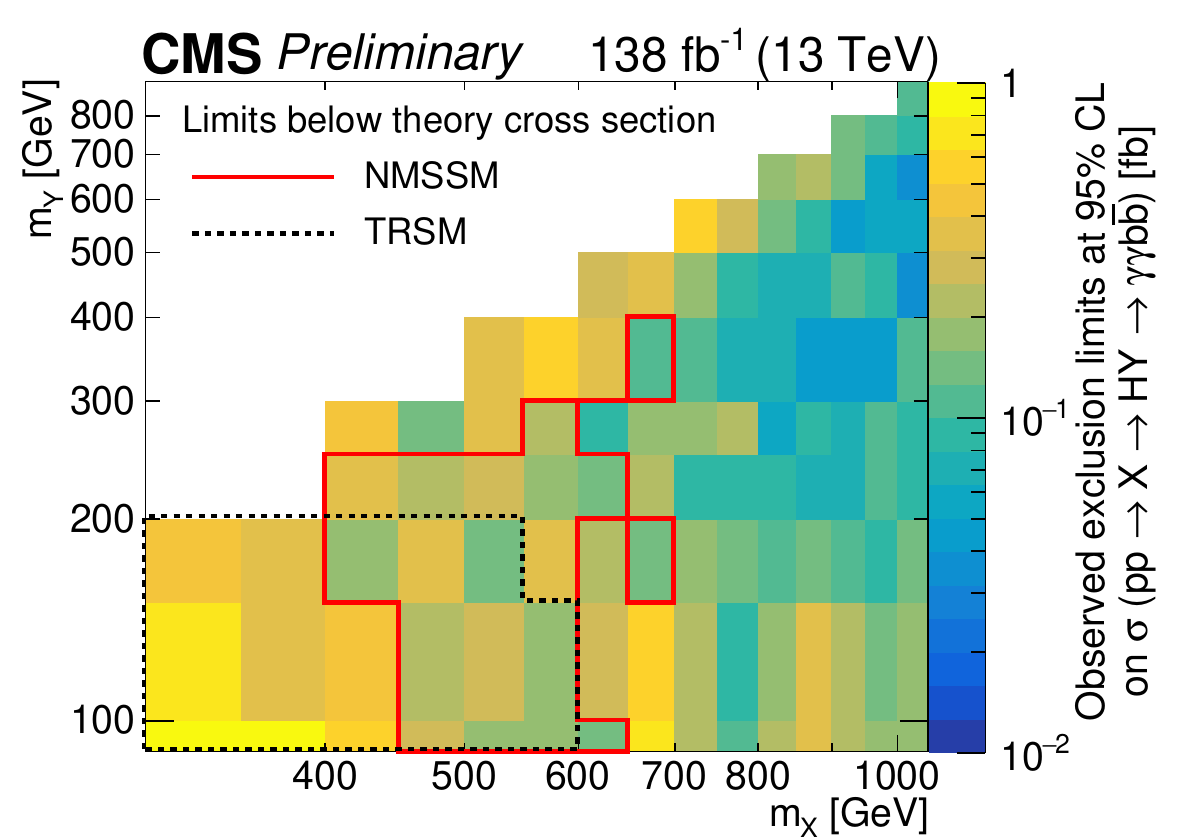}
\caption{\label{fig:cmsbbgaga} Expected {\sl (left)} and observed {\sl (right)} $95\%$ confidence limits for the $p\,p\,\rightarrow\,h_3\,\rightarrow\,h_2\,h_1$ search, with subsequent decays into $b\bar{b}\gamma\gamma$. Depending on the model, maximal mass regions up to $m_3\,\sim\,\,800\,\GeV,\;m_2\,\sim\,\,400\,\GeV$ can be excluded.}
\end{center}
\end{figure}
\end{center}

Two experimental searches have by now made use of the predictions obtained within the TRSM to interpret regions in parameter space that are excluded: a CMS search for asymmetric production and subsequent decay into $b\bar{b}b\bar{b}$  final states \cite{CMS:2022suh}, as well as $b\bar{b}\gamma\gamma$ in \cite{cms}. Maximal rates for these within the TRSM are documented in \cite{reptr,trsmbbgaga}. Figures \ref{fig:cmsres} (taken from \cite{CMS:2022suh}) and \ref{fig:cmsbbgaga} (taken from \cite{cms}) show the expected and observed limits in these searches for the TRSM and NMSSM \cite{Ellwanger:2022jtd}.
\section{Processes at Higgs factories}
In the European Strategy report \cite{EuropeanStrategyforParticlePhysicsPreparatoryGroup:2019qin,cern}, Higgs factories were identified as one of the high priority projects after the HL-LHC. At such machines, new physics scalar states could also be produced in the mass range up to $\sim\,160\,\GeV$ depending on the collider process.

\begin{center}
\begin{figure}[tbh]
\begin{center}
\includegraphics[width=0.45\textwidth]{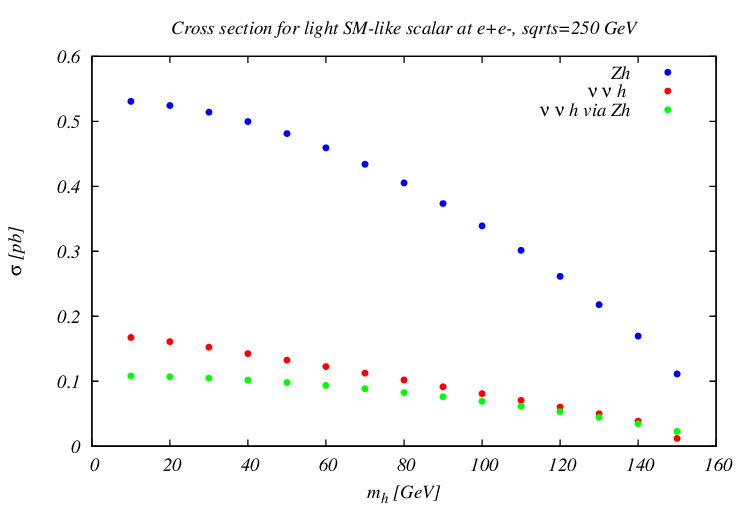}
\caption{\label{fig:prod250} Leading-order production cross sections for $Z\,h$ and $h\,\nu_\ell\,\bar{\nu}_\ell$ production at an $e^+\,e^-$ collider with a com energy of 250 \GeV~  using Madgraph5 for an SM-like scalar $h$. Shown is also the contribution of $Z\,h$ to $\nu_\ell\,\bar{\nu}_\ell\,h$ using a factorized approach for the Z decay. Update of plot in \cite{Robens:2022zgk}, first presented in \cite{Robens:2022uis}.}
\end{center}
\end{figure}
\end{center}

In general, additional light scalar states can be singly produced e.g. via so-called scalar-strahlung $e^+e^-\,\rightarrow\,Z\,h$, or in the VBF-type topologies with $e^+\,e^-\,\rightarrow\,v_e\,\bar{\nu}_e\,h$. As we cannot impose a direct cut on the rapidity of the neutrinos, also scalar strahlung can contribute to the latter final state if $Z\,\rightarrow\,\nu_e\,\bar{\nu}_e$.  Leading-order predictions for $Zh$ production at $e^+e^-$ colliders for low mass scalars which are Standard Model (SM)-like, using Madgraph5 \cite{Alwall:2011uj}, are shown in figure \ref{fig:prod250} for a center-of-mass energy of 250 \GeV.   While for lower masses VBF production still plays a role, for higher scalar masses the dominant contribution stems from $Z\,h$ production.

\begin{center}
\begin{figure}[tbh]
\begin{center}
\includegraphics[width=0.45\textwidth]{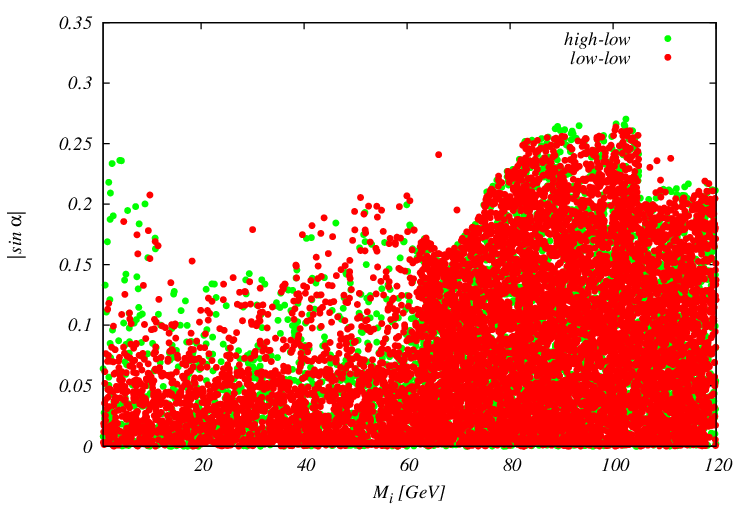}
\includegraphics[width=0.45\textwidth]{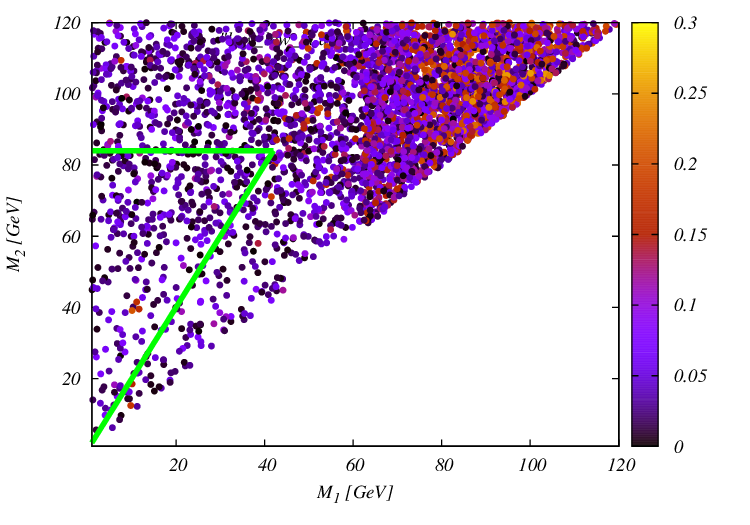}
\caption{\label{fig:eespace}  Available parameter space in the TRSM, with one (high-low) or two (low-low) masses lighter than 125 \GeV. {\sl Left}: light scalar mass and mixing angle, with $\sin\al\,=\,0$ corresponding to complete decoupling. {\sl Right:} available parameter space in the $\lb m_{h_1},\,m_{h_2}\rb$ plane, with color coding denoting the rescaling parameter $\sin\al$ for the lighter scalar $h_1$. Within the green triangle, $h_{125}\,\rightarrow\,h_2 h_1\,\rightarrow\,h_1\,h_1\,h_1$ decays are kinematically allowed. Taken from \cite{Robens:2022zgk}.}
\end{center}
\end{figure}
\end{center}

\begin{center}
\begin{figure}[tbh]
\begin{center}
\includegraphics[width=0.45\textwidth]{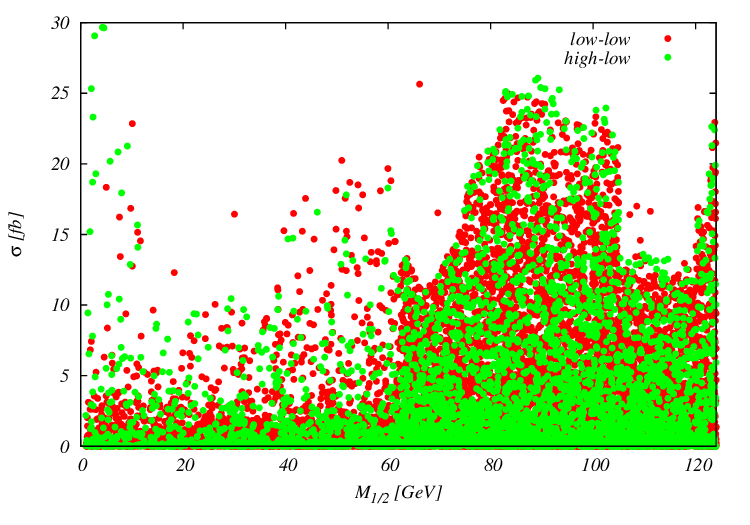}
\includegraphics[width=0.45\textwidth]{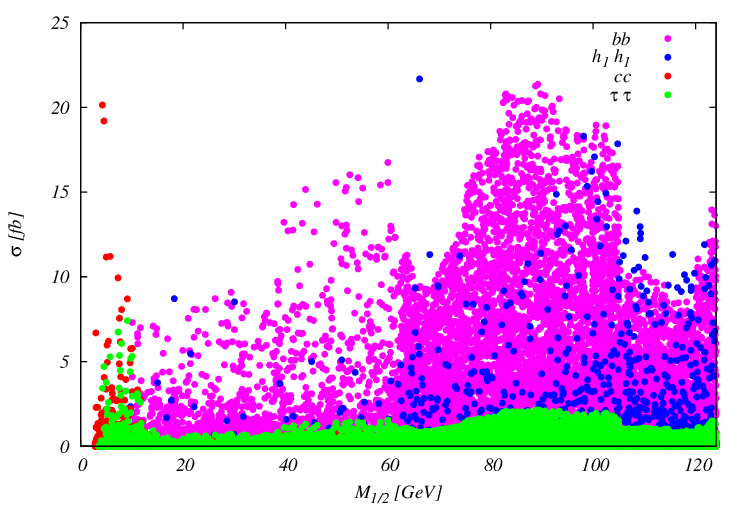}
\caption{\label{fig:trsm_conv} Points in the TRSM taken from figure \ref{fig:eespace}, and convoluted with production cross sections at a 250 \GeV~ Higgs factory {\sl (left)}, and additionally with branching ratios of dominant final states {\sl (right)}. Update of a figure first presented in \cite{Robens:2022nnw}.}
\end{center}
\end{figure}
\end{center}

It is then interesting to see which parameter range is still available for such light scalar states in the TRSM, where now $\sin\al$ corresponds to the rescaling angle with respect to the coupling of a SM-like Higgs at a certain mass $M_i$. The allowed parameter space is shown in figure \ref{fig:eespace}. Results after convolution with production cross sections and decay rates are show in figure \ref{fig:trsm_conv}. We see that total production rates can be in the $20\,-\,30\,\fb$ range, while dominant decay rates are into $b\,\bar{b}$ and $h_1\,h_1$ final states; the latter would lead to $Z\,h_1\,h_1$ with typical decays to $Z\,b\,\bar{b}\,b\,\bar{b}$ final states, a signature that so far has not been explored yet. We also observe $c\,\bar{c}$ and $\tau^+\,\tau^-$ final states in some regions of the parameter space. Ref. \cite{filip_here} shows some preliminary studies of discovery prospects.

\section{Conclusions}
The TRSM is a new physics model that, with a small number of additional new physics parameters, allows for novel final states, in particular asymmetric scalar production and decays. I briefly summarized the current status of benchmark planes originally proposed in \cite{Robens:2019kga} as well as possible rates for this model at future facilities, such as Higgs factories.  UFO file for the model and maximal cross section values are available upon request.

\section*{Acknowledgements}
The work was partially supported by the National Science Centre (Poland) under the OPUS research project no. 2021/43/B/ST2/01778.


\end{document}